\documentclass{iopjournal}

\begin{document}

\articletype{Paper} 

\title{Towards real-time ion range verification via hybrid Compton-PET 3D imaging at isochronous cyclotrons}

\author{
Javier Balibrea-Correa$^{1,*}$\orcid{0000-0002-8404-3256}, 
V. Babiano-Suarez$^1$\orcid{0000-0000-0000-0000},
J. Lerendegui-Marco$^{1}$\orcid{0000-0001-8358-9217},
P. Torres-Sánchez$^{1}$\orcid{0000-0000-0000-0000},
I. Ladarescu$^{1}$\orcid{0000-0003-2405-281X},
M. Pall\`as$^{2,3,4}$\orcid{0000-0001-6137-0812},
B. Brusasco$^{2}$\orcid{0000-0000-0000-0000},
J. Wulff$^{5}$\orcid{0000-0001-8260-3523},
C. Bäumer$^{5}$\orcid{0000-0003-0512-1146},
A. Tarifeño-Saldivia$^{1}$\orcid{0000-0001-8552-5030},
and C. Domingo-Pardo$^{1}$\orcid{0000-0002-2915-5466}}

\affil{$^1$Instituto de Física Corpuscular (CSIC-University of Valencia), Paterna, Spain}

\affil{$^2$Universitat Politècnica de Catalunya, Barcelona, Spain. }

\affil{$^3$GSI Helmholtzzentrum f\"ur Schwerionenforschung, Darmstadt, Germany. }

\affil{$^4$Institut f\"ur Kernphysik, Technische Universit\"at Darmstadt, Darmstadt, Germany. }

\affil{$^5$West German Proton Therapy Centre, Essen, Germany. } \newline
\affil{$^*$Author to whom any correspondence should be addressed.}

\email{javier.balibrea@ific.uv.es}


\begin{abstract}
Range uncertainties remain a major limitation to the full clinical exploitation of proton therapy, motivating the development of reliable in-vivo range-verification techniques. This work presents an experimental investigation of a hybrid Prompt-Gamma Imaging (PGI)--Positron-Emission Tomography (PET) concept at the isochronous cyclotron of the West German Proton Therapy Centre, under proton-beam delivery conditions representative of clinical pencil-beam-scanning treatments.

Because this accelerator intrinsically delivers a quasi-continuous beam, a pulsed time structure was imposed by alternating shoot-and-step periods, mimicking the beam interruptions associated with clinical energy-layer switching. This enabled PGI acquisition during irradiation and PET and delayed Compton imaging during step intervals within the same irradiation sequence, reproducing the temporal separation naturally available in synchrocyclotron-based systems. Four two-plane Compton-camera modules were arranged in a co-planar, cross-shaped geometry around the beam isocenter.

Polyethylene phantoms were irradiated with 100~MeV protons at accurately defined positions along the beam axis. Their positions were estimated from one-dimensional profiles of the reconstructed distributions using supervised machine-learning models trained exclusively on Monte Carlo simulations of the experimental setup.

PET provided the best performance, with a root-mean-square deviation of approximately 0.8~mm, followed by PGI with 1.6~mm. Delayed Compton imaging based on e$^{+}$ annihilation photons and $^{10}$C$^{*}$ decays yielded deviations of 2.4~mm and 3.8~mm, respectively. A dedicated reduced-statistics study further showed that PET retained good position-estimation performance down to 1\% of the original event statistics, whereas PGI became progressively limited by statistical fluctuations.

These results experimentally demonstrate the complementary strengths of the hybrid approach, combining the prompt response of PGI with the higher position-estimation accuracy and statistical robustness of PET, and support its further development towards millimetre-scale proton-range verification under clinically relevant irradiation conditions.
\end{abstract}

\section{Introduction}
\label{sec:introduction}

Proton therapy has become an established treatment option for a range of solid tumours, particularly in pediatric patients and for lesions located close to radiosensitive organs~\cite{Durante:2016,Durante:2017}. Its principal physical advantage over conventional radiotherapy arises from the characteristic depth--dose profile of charged particles, which deposit most of their energy within a narrow region near the end of their trajectory, known as the Bragg peak. This allows the dose delivered to proximal healthy tissue to be reduced while providing negligible irradiation beyond the target volume~\cite{Paganetti:2012}. Consequently, proton therapy enables highly conformal dose distributions and may reduce the risk of secondary malignancies and other long-term complications associated with conventional radiotherapy~\cite{Knopf:13,ParodiPolf:2018}.

Despite these intrinsic advantages, the full clinical potential of proton therapy remains limited by uncertainties in the particle range within the patient. Treatment planning traditionally relies on converting X-ray computed-tomography Hounsfield units into proton stopping powers using semi-empirical calibration curves, introducing systematic range uncertainties of approximately 1--3\%~\cite{Yang:2012,Moyers:2010}. Historically, these uncertainties have been accommodated through conservative safety margins of up to 3.5\% + 3~mm~\cite{Paganetti:2012}, resulting in the irradiation of additional healthy tissue beyond the intended target. The introduction of dual-energy CT into treatment planning has substantially improved stopping-power estimation and enabled range-related margins to be reduced towards $\sim$2\% in clinical practice~\cite{WOHLFAHRT17,Peterson24,Schwengfelder26}. Nevertheless, residual uncertainties arising from image-to-stopping-power conversion, anatomical variations and treatment delivery remain. The development of real-time, \textit{in vivo} techniques capable of directly verifying the proton range with sub-millimetric accuracy would therefore represent a major advance, potentially allowing further reduction of safety margins and expansion of the clinical indications of hadron therapy~\cite{Knopf:13}. Several approaches are currently under investigation, including proton-acoustic techniques~\cite{Jones:18}, non-imaging prompt-gamma count-rate and timing monitoring~\cite{Hueso:20}, and detection of secondary charged particles produced during irradiation~\cite{HUANG:24}, in addition to the imaging-based techniques discussed below.

Two $\gamma$-ray imaging techniques have been extensively investigated for in-beam ion-range verification: Positron Emission Tomography (PET) and Prompt-Gamma Imaging (PGI). PET-based range monitoring exploits the $\beta^{+}$-emitting isotopes produced along the proton trajectory through nuclear reactions with tissue nuclei, predominantly the comparatively long-lived isotopes $^{11}$C, $^{13}$N, and $^{15}$O~\cite{Enghardt:04,PARODI:2007}. The coincident detection of the two 511~keV annihilation photons enables a high-resolution spatial reconstruction of the activated volume, from which the distal edge of the dose distribution can be inferred. However, in-beam PET presents several challenges, including biological washout of the activated isotopes, relatively low signal-to-background ratios, pair-production backgrounds, and the need for non-standard detector geometries compatible with the treatment environment~\cite{Shakirin:2011,Ferrero:2018,Bisogni:2016}. Furthermore, the intrinsically broad spatial distribution of the $\beta^{+}$ emitters relative to the Bragg peak introduces an additional source of systematic uncertainty in the retrieved particle range~\cite{Moteabbed:2011}.

PGI, in contrast, exploits the quasi-instantaneous emission of high-energy secondary $\gamma$-rays, extending up to approximately 5--7~MeV, produced in prompt nuclear reactions along the ion trajectory~\cite{Stichelbaut:03,Min:06}. In addition to a continuous $\gamma$-ray energy emission, discrete characteristic $\gamma$-ray lines arise from inelastic nuclear reactions between the incident protons and light nuclei present in tissue, most prominently $^{12}$C, $^{14}$N, and $^{16}$O, with the 4.44~MeV ($^{12}$C) and 6.13~MeV ($^{16}$O) transitions among the most intense~\cite{Jeyasugiththan:21}. These characteristic lines encode direct information on the elemental composition of the irradiated medium and can, in principle, be isolated from the underlying continuum through dedicated spectral-unfolding techniques~\cite{HOSOBUCHI:23}, providing a complementary, composition-sensitive source of range information; the present work, however, exploits only the broadband continuum PGI signal and does not make use of these discrete lines.

The strong spatial and temporal correlation between this prompt $\gamma$-ray emission and the Bragg peak makes PGI particularly attractive for real-time range monitoring. A mechanically collimated slit camera has already been clinically validated with satisfactory results~\cite{Smeets:2012,Perali:2014}; however, its one-dimensional sensitivity and intrinsically low detection efficiency, of the order of $10^{-5}$, limit the accuracy. Electronic collimation can reconstruct the photon origin from the kinematics of a Compton-scattering interaction rather than from a mechanical aperture overcoming this efficiency limitation and provide two-dimensional spatial reconstruction~\cite{Babiano:20}. Nevertheless, it introduces additional challenges associated with the high instantaneous counting rates, intense neutron-induced backgrounds, and predominance of random coincidences under clinical Pencil-Beam Scanning (PBS) conditions~\cite{Krimmer:18,Testa:08,PAUSCH:2020}.

The complementary strengths and limitations of PET and PGI motivated the hybrid detection concept first proposed by Parodi in 2016~\cite{Parodi:16}. In this framework, PGI is based on Compton imaging performed during the delivery of the beam, or shoot periods, using the prompt $\gamma$-ray emission, whereas PET imaging exploits the $\beta^{+}$ activity recorded during the waiting time between successive proton deliveries, referred to as step intervals. Additionally, Compton imaging can be used to target e$^{+}$ annihilation $\gamma$-rays and $^{10}$C$^{*}$ decay transitions. As suggested by Lang~\textit{et al.}~\cite{Lang:14}, this concept can be naturally implemented using an array of Compton cameras arranged in a geometry that supports both imaging modalities. The hybrid approach is expected to provide complementary spatial and temporal information: PGI offers a direct and quasi-instantaneous signature of the Bragg-peak position with a high event yield, whereas PET provides high intrinsic spatial resolution and tomographic information related to the $\beta^{+}$-activity distribution.

The feasibility of this hybrid concept has been progressively demonstrated through a series of Monte Carlo (MC) simulations and experimental campaigns, starting with a detailed MC feasibility study of the PGI component~\cite{Lerendegui:2022}. An initial proof-of-concept PGI-PET experiment was conducted at the 18~MeV cyclotron radiobiology beam line of the Centro Nacional de Aceleradores in Seville using two Compton cameras in a face-to-face configuration~\cite{Balibrea22a,Balibrea22b}. These measurements demonstrated submillimetric PET position accuracy together with simultaneous PGI reconstruction, establishing the basis for subsequent investigations at clinically relevant beam energies. A later preclinical campaign at the Heidelberg Ion Therapy Centre (HIT) extended the study to proton, helium, and carbon-ion beams with energies representative of clinical treatments, ranging from 55 to 275~MeV depending on the incident particle. These measurements revealed a pronounced dependence of the performance of both imaging modalities on the beam energy and ion species and highlighted the challenges encountered by PGI for heavier ions at high energies~\cite{Balibrea:2025a}.

In this work, we present a comprehensive experimental investigation of the hybrid PGI--PET technique performed at the West German Proton Therapy Centre (WPE) in Essen. The experiment was designed to move this research line beyond the proof-of-concept and exploratory stages established at CNA and HIT towards a rigorous benchmark under clinically representative irradiation conditions. This advance is defined by three main developments.

First, the measurements were performed with an IBA Proteus Plus isochronous cyclotron under conditions representative of clinical pencil-beam-scanning treatments. Cyclotron-based systems are the accelerator technology most widely deployed in clinical proton-therapy facilities worldwide, in contrast to the synchrotron-based beam delivery used in the previous campaign at HIT.

Second, the detector geometry was substantially improved. At HIT, the four Compton cameras operated as two independent face-to-face pairs along the beam axis, a configuration that was not optimal in terms of detection efficiency, particularly for PGI sensitivity (see Fig.1 in \cite{Balibrea:2025a}). At WPE, the four Compton-camera arms were instead arranged in a co-planar cross-shaped configuration within a single transverse plane (see Fig.\ref{fig:setup} below), following the geometry proposed in Refs.~\cite{Parodi:16,Lerendegui:2022}, which is expected to increase the PGI detection efficiency by about a factor of two.

Third, the Gaussian-fit peak-centroiding approach used for position determination in the earlier CNA and HIT campaigns was replaced by a dedicated quantitative methodology based on machine learning (ML). The models were trained exclusively on MC simulations of the experimental setup and then applied directly to the measured data.

Together, these developments enable, for the first time, a quantitative comparison of the range-verification performance achieved with PGI, PET, and Compton imaging under realistic clinical beam-delivery conditions, resolved across different levels of event statistics. The WPE campaign therefore establishes a rigorous experimental benchmark for the hybrid concept rather than a further proof-of-concept demonstration.

The sensitivity, spatial resolution, and range-verification capabilities of the hybrid system are subsequently evaluated. A preliminary account of this work was reported in Ref.~\cite{Balibrea:2025b}. The remainder of the paper is organised as follows. Section~\ref{sec:setup} describes the experimental setup and beam-delivery conditions. Section~\ref{sec:analysis} presents the image-reconstruction procedures and the ML-based methodology used for proton-range verification. The experimental results are presented in Section~\ref{sec:full-statistics} together with a statistics sensitivity study in Section~\ref{sec:reduced-statistics}. The analysis is followed by the discussion and conclusions in Sections~\ref{sec:discussion} and~\ref{sec:conclusions}, respectively.

\section{Methodology}

The hybrid PGI--PET concept relies on the intrinsic separation between beam delivery and step time  intervals to acquire, respectively, prompt-gamma and PET/Compton imaging data within the same irradiation sequence. Commercial isochronous cyclotrons, such as the IBA Proteus Plus system installed at WPE and employed in this work, intrinsically deliver a quasi-continuous-wave proton beam rather than a pulsed one. For the purposes of the present study, a pulsed time structure was therefore imposed on the beam delivery (see Section~\ref{sec:setup}), enabling the acquisition of PGI and PET/Compton data within the same irradiation sequence. Qualitatively, the resulting alternation between shoot and step periods is comparable to the beam interruptions that occur naturally during energy-layer switching in clinical pencil-beam-scanning delivery, rather than constituting a delivery scheme alien to clinical practice.

The primary objective of the experimental campaign was to determine the position of polyethylene (PE) phantoms along the beam axis using four independent imaging observables: reconstructed PGI distributions, PET distributions, Compton images of 511~keV e$^{+}$-annihilation photons, and Compton images of the 718~keV $\gamma$-rays emitted following $^{10}$C$^*$ decay.


\subsection{Experimental setup}
\label{sec:setup}

The proton beam at WPE was delivered to the treatment room in PBS mode at an energy of 100~MeV and with a nominal spot diameter of 17.6~mm full-width-at-half-maximum. A minimum clinical beam current of 0.3~nA was maintained throughout the experiment to keep at minimum the dead-time in the acquisition system. A pulsed beam time structure was employed, consisting of beam-on intervals of 400--800~ms followed by beam-off periods of 1--2~s, comparable in nature to the beam interruptions associated with energy-layer switching during clinical pencil-beam-scanning delivery. About 40 proton spots were delivered for each individual irradiation, each containing $1 \times 10^{9}$ protons, providing sufficient statistics for testing the accuracy of all imaging observables. PGI data were acquired exclusively during the shoot periods, whereas the step intervals were used to acquire data for PET and Compton imaging of $e^{+}$ and $^{10}$C$^{*}$.

\begin{figure}
\centering
\includegraphics[width=0.8\textwidth]{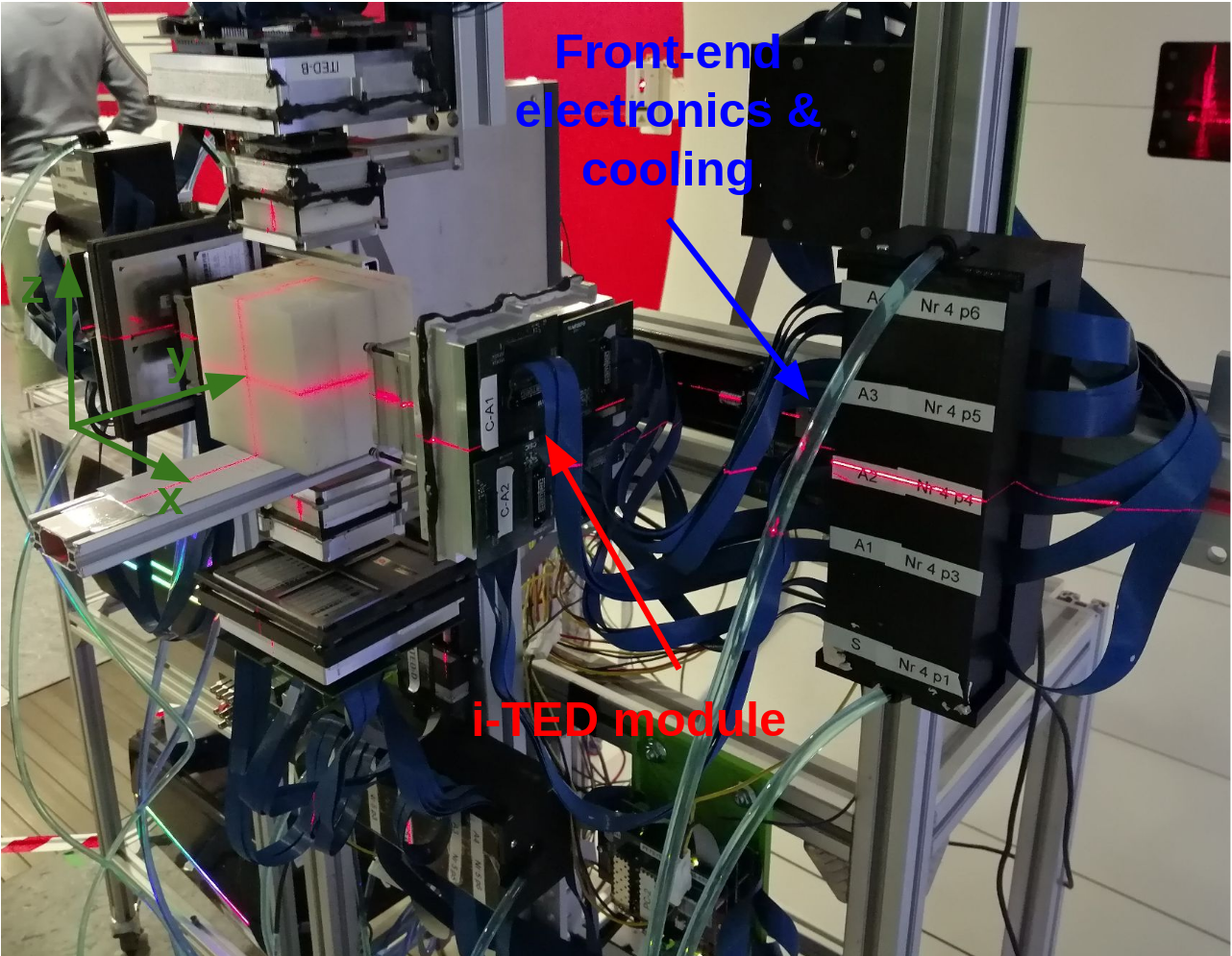}
\caption{Photograph of the hybrid PGI--PET detection system used during the experimental campaign at WPE. The setup consisted of four Compton cameras arranged in a co-planar, cross-shaped geometry. One of the i-TED Compton cameras is shown together with the associated front-end electronics and cooling system. For completeness, the global coordinate system used for image reconstruction is also indicated.}
\label{fig:setup}
\end{figure}

The hybrid PGI--PET detection system comprised four two-plane Compton cameras with an oversize absorber plane, hereafter referred to as i-TED modules~\cite{DOMINGOPARDO:2016,Babiano:20}. As shown in Fig.~\ref{fig:setup}, the modules were arranged in a co-planar, cross-shaped configuration around the proton-beam isocenter and positioned symmetrically with respect to the beam axis. The distance between the front faces of two opposing modules was 22.3~cm.

For Compton imaging, a valid event was defined as the time-coincident detection of two interactions within the same module. The incident photon first undergoes Compton scattering in the scatter plane, depositing part of its energy through the recoil electron, and the scattered photon is subsequently absorbed in one of the four crystals forming the absorber plane.

The measured interaction positions and deposited energies were then combined through the Compton-scattering kinematics to reconstruct the corresponding Compton cone. Its apex is given by the interaction point in the scatter plane, its axis by the line connecting the scatter- and absorber-plane interaction points, and its half-opening angle by the Compton-scattering relation. The original photon-emission point is constrained to lie on the surface of this cone~\cite{Kim:24}.

PET imaging, in contrast, relied on the coincident detection of e$^{+}$ annihilation photons in opposing i-TED modules. A coincidence-time window of 10~ns was applied in both modalities. The cross-shaped arrangement of the four modules allowed all detector combinations to contribute to the reconstruction, enabling three-dimensional imaging with both Compton and PET techniques.

Compared with the twofold front-to-front geometry of four i-TED modules employed in the previous experimental campaign at HIT~\cite{Balibrea:2025a}, in which the modules operated as two independent opposing pairs, the cross-shaped arrangement adopted here combines the information of all four modules simultaneously within a single transverse plane, substantially increasing the detection efficiency and available statistics for PGI. This improvement came at the expense of a more restricted PET field of view (FoV) along the beam axis, primarily determined by the dimensions of the absorber planes, $120 \times 120$~mm$^2$. In contrast, the PGI FoV extended over a considerably larger region of approximately $400 \times 400$~mm$^2$, centered on the beam axis. It should be noted, however, that both the Compton reconstruction efficiency and angular resolution deteriorate significantly for source angles exceeding $45^{\circ}$ with respect to the camera axis~\cite{Balibrea22b}.

During the experiment, 10$\times$10$\times$7.5~cm$^3$ PE phantoms were irradiated at a series of accurately defined positions along the beam axis, spanning from -40 to +40~mm relative to the geometric center of the experimental setup. The investigated positions were -40, -30, -25, -20, -15, -10, -8, -6, -4, -2, 0, 2, 4, 6, 8, 10, 15, 20, 25, 30, and 40~mm. This coordinate convention is used throughout the remainder of the paper.

After each irradiation, the PE phantom was replaced with a fresh one to prevent residual activity from relatively long-lived $\beta^{+}$ emitters produced during the preceding irradiation from contaminating subsequent measurements.

The individual LaCl$_3$(Ce) detectors were calibrated using standard $^{152}$Eu and $^{22}$Na radioactive sources, together with the intrinsic $\alpha$-particle background present in the scintillation crystals~\cite{Balibrea:2025a,Babiano:20}. The relative positions of the four i-TED modules were determined by placing a quasi-point-like $^{22}$Na source at accurately known positions along the beam axis within the PET FoV. 

\subsection{Image reconstruction and machine-learning methodology}
\label{sec:analysis}

Dedicated GPU-based implementations were employed for each imaging modality. PET images were reconstructed using the Stochastic Origin Ensemble (SOE) algorithm~\cite{Andreyev09}, whereas Compton images were obtained with a Maximum-Likelihood List-Mode (MLLM) algorithm incorporating median-filter regularization~\cite{Sakai20}.

The SOE algorithm reconstructs the source distribution directly from the list-mode coincidence data, without requiring an explicit system matrix. For each recorded coincidence, a candidate emission point is randomly sampled along the corresponding line of response; over successive iterations, these candidate points are stochastically redistributed according to their local density relative to neighboring events, so that the resulting ensemble of points converges towards an estimate of the true activity distribution.

The MLLM algorithm follows an analogous list-mode approach for Compton imaging. Each valid Compton event constrains the photon origin to the surface of a cone, and the voxelised image is updated iteratively, event by event, through a maximum-likelihood expectation-maximization scheme that compares the current image estimate with the geometrical constraint imposed by each cone. A median-filter regularization step is applied between iterations to suppress the streak-like artifacts characteristic of cone back-projection and to improve convergence.

PGI images were reconstructed using all valid Compton events, applying an energy-selection of deposited energy larger than 250~keV. For the step acquisition period, add-back deposited-energy windows of $511 \pm 50$~keV and $718 \pm 50$~keV were applied to reconstruct the positron-annihilation and $^{10}$C$^{*}$-related activity distributions, respectively. For PET reconstruction, an additional energy-selection criterion was imposed, requiring each of the two coincident photons to deposit an energy within the $511 \pm 50$~keV window in the corresponding detector to suppress the contribution from random coincidences.

\begin{figure}[htbp]
\centering
\includegraphics[width=0.8\linewidth]{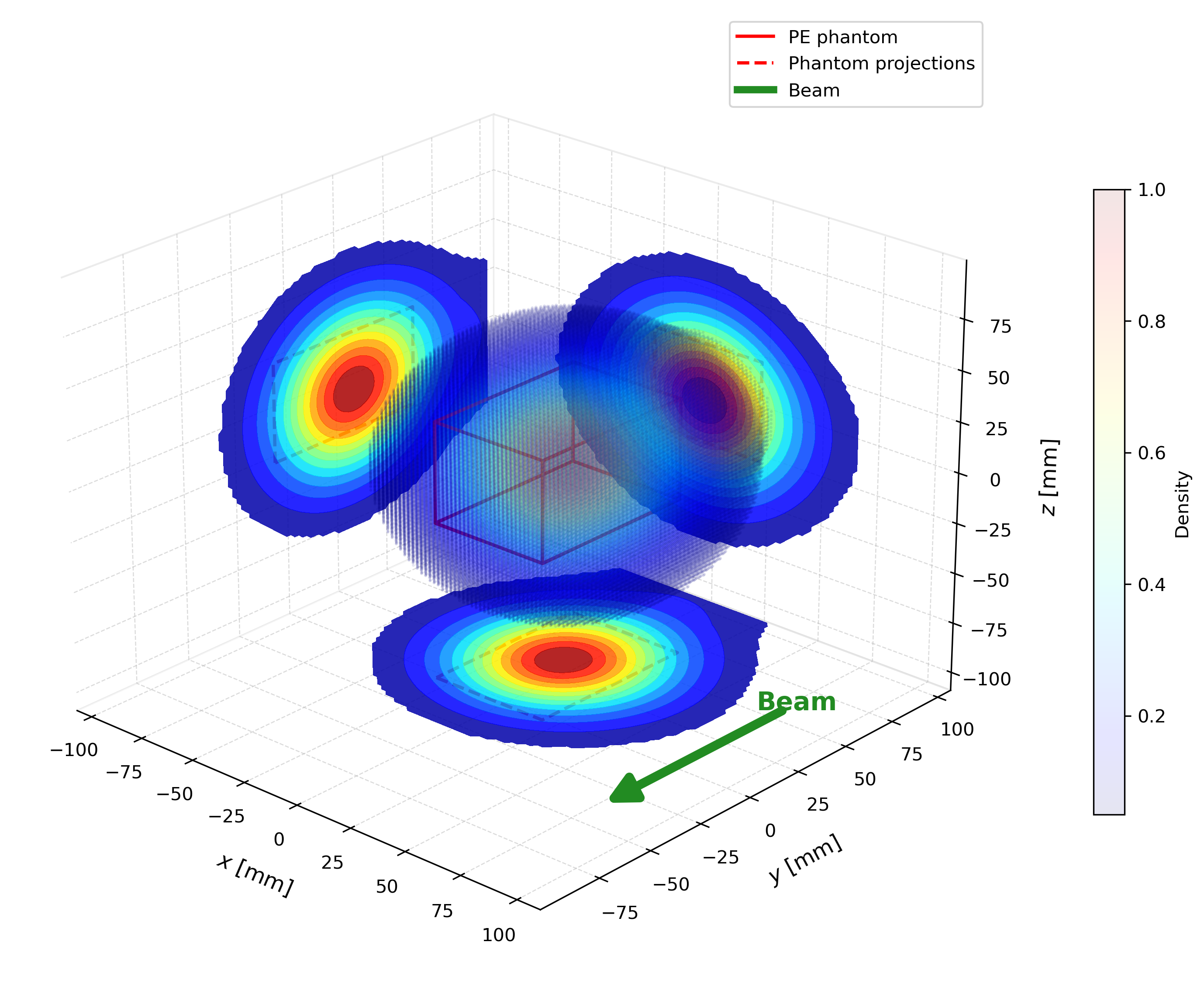}
\caption{Example of full three-dimensional reconstructed PGI distribution for one of the PE-phantom irradiation. See the text for further details.}
\label{fig:compton3d_distribution}
\end{figure}

\begin{figure}[htbp]
\centering
\includegraphics[width=0.8\linewidth]{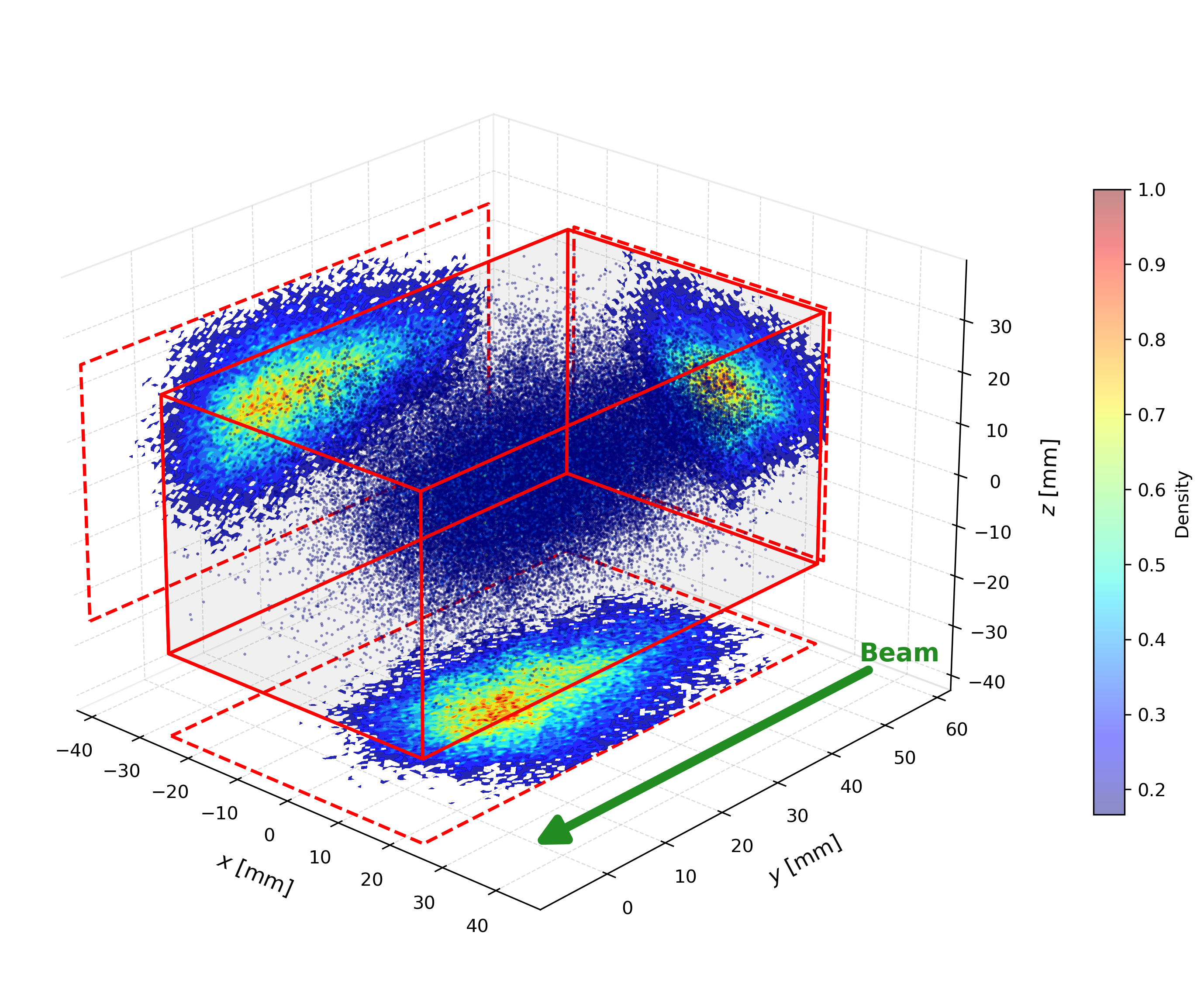}
\caption{Example of full three-dimensional activity distribution reconstructed using PET data. See the text for further details.}
\label{fig:pet3d_distribution}
\end{figure}

Examples of the reconstructed PGI and PET distributions are shown in Figs.~\ref{fig:compton3d_distribution} and~\ref{fig:pet3d_distribution}, respectively. In both cases, the reconstructed activity distribution is represented as a three-dimensional point cloud in the $(x,y,z)$ coordinate system of the experimental setup, with the color scale normalized from low (blue) to high (red) activity density. The corresponding projections onto the three orthogonal planes are displayed on the back, side, and bottom planes of the bounding box.

The true position and dimensions of the PE phantom are indicated by a solid red cube, while dashed red outlines represent its projections onto the three planes. The proton-beam direction is indicated by a green arrow. This representation enables a direct comparison between the reconstructed activity distribution and the known phantom geometry. In addition, the figures illustrate the markedly different fields of view obtained for PET and PGI, arising from the intrinsic characteristics and geometrical constraints of the two imaging modalities.

Although the reconstructed three-dimensional distributions provide qualitative information on the phantom location, the relationship between the reconstructed signal profile and the true phantom position is strongly non-linear. A supervised ML approach was therefore adopted to estimate the phantom position along the beam axis.

Several supervised regression algorithms were evaluated, including \emph{k}-nearest neighbours, random forests, gradient-boosted decision trees, and support-vector machine regression (SVM). Their description and implementation can be found in Ref.~\cite{scikit-learn}. Among these methods, SVM with a linear kernel provided the best overall performance for estimating the PE-phantom position and was therefore adopted for the final analysis. Although both PET and PGI images were reconstructed in three dimensions, only the one-dimensional profiles obtained by projecting the reconstructed distributions along the beam axis were used as input to the regression model. This approach was decided to decrease the computational cost as the remaining information about the other axis did not play a major role.

Following a strategy conceptually analogous to that used in treatment planning based on MC~\cite{Paganetti:2012,Böhlen2013A,Fracchiolla2021}, the ML models were trained exclusively on realistic MC simulations of the experimental setup. A detailed model of the experiment, including the four i-TED modules and the PE phantom, was implemented in a dedicated C++ application based on the Geant4 toolkit~\cite{ALLISON:2016}, following the modeling approach previously adopted for the CNA and HIT campaigns~\cite{Balibrea22a,Balibrea:2025a}. The simulation employed the standard electromagnetic physics option 4 together with the appropriate hadronic and radioactive-decay models for reproducing prompt-$\gamma$ emission and $\beta^{+}$-emitting nuclei distribution~\cite{Verburg:2012}. 

 A total of 81 phantom positions were simulated, uniformly spanning the range from $-40$ to $+40$~mm in 1~mm steps. For each position, $10^{10}$ incident protons were simulated through the PE phantom. Both prompt and delayed events, corresponding to the PGI and PET acquisition periods, respectively, were recorded. The experimental energy and position resolutions determined during detector characterization were subsequently applied to the simulated events prior to image reconstruction. The resulting datasets were then processed using the same reconstruction procedures as the experimental data, thereby reproducing the measurement conditions as closely as possible. 
Once trained on the MC dataset, the ML model was applied directly to the corresponding experimental one-dimensional profiles to obtain an estimate of the phantom position along the beam axis.

\begin{figure}[htb!]
\centering
\begin{tabular}{c c}
\includegraphics[width=0.5\columnwidth]{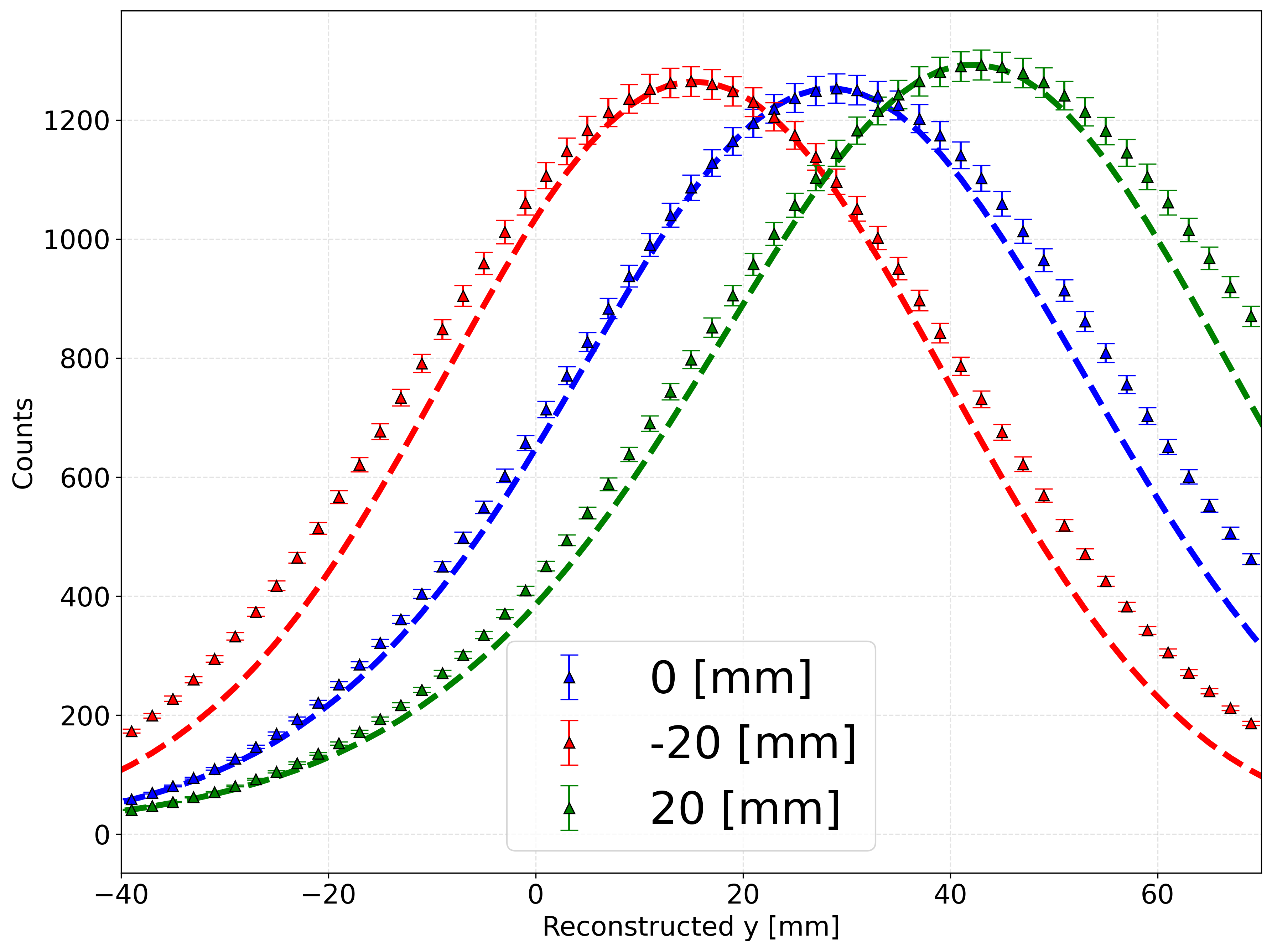} &
\includegraphics[width=0.5\columnwidth]{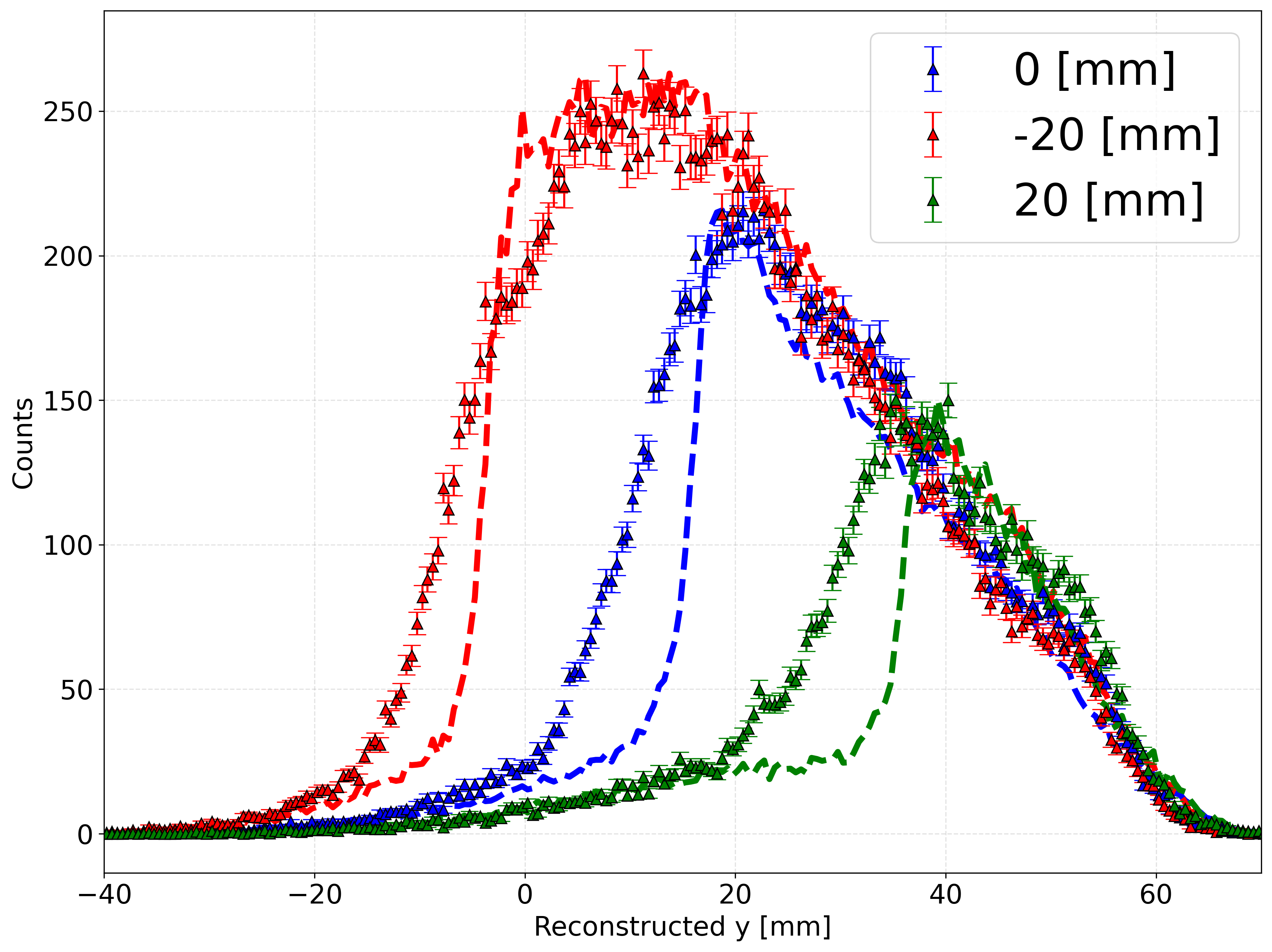}
\end{tabular}
\caption{Representative experimental one-dimensional activity distributions projected along the beam axis for PGI (left) and PET (right). The corresponding MC distributions are shown using dashed lines of the same colour.}
\label{fig:FitDistributions}
\end{figure}

As an example, representative experimental and matching simulated profiles are shown in Fig.~\ref{fig:FitDistributions} for PGI (left panel) and PET (right panel). Overall, good agreement was observed between the measured and simulated distributions for both modalities. For PET, however, the simulated profiles exhibit a flatter, plateau-like top and a steeper rising edge than the smoother, more rounded experimental distributions, indicating a small but systematic shape mismatch between experimental and MC simulation. This discrepancy was attributed mainly to the simplified description of the $\beta^{+}$-emitter composition in the simulations, in which all produced isotopes were included without tuning their relative production yields to reproduce the experimentally observed isotope mixture. Since PGI relies on the quasi-instantaneous emission of prompt $\gamma$-rays rather than on the decay of activated isotopes, it is not subject to this particular source of uncertainty, consistent with the closer, near-featureless agreement observed between the experimental and simulated PGI profiles.

The small residual differences between the experimental and MC-reconstructed profiles resulted in a compression of the phantom positions estimated from the experimental data. To compensate for this systematic effect, a linear calibration was applied to the ML-estimated position, $x$, yielding the corrected position, $y$, according to

\begin{equation}
y=\frac{x-b}{a},
\end{equation}

where $a$ and $b$ are the slope and intercept, respectively, obtained from a linear fit relating the ML-estimated positions to the corresponding known experimental phantom positions. The uncertainties associated with the calibration parameters were propagated to the corrected positions using standard error-propagation methods. In addition, a systematic uncertainty was assigned from the standard deviation of the calibration points around the fitted linear relation, accounting for the residual dispersion not described by the calibration.

It should be noted that $a$ and $b$ are obtained from the same set of phantom positions subsequently used to evaluate the position residuals. However, since the calibration involves only two global parameters common to all positions, it can only correct for a uniform offset and slope in the ML-estimated positions; it cannot reproduce or artificially suppress any position-dependent or point-to-point scatter beyond this simple linear relation. The residual dispersion reported in this work therefore reflects a genuine limitation of the position-estimation method, rather than an artifact of the calibration procedure.

\section{Results}
\label{sec:results}

This section presents the position-estimation results obtained for the four imaging observables described in Section~\ref{sec:analysis}. Section~\ref{sec:full-statistics} reports the performance achieved using the full experimental statistics (40$\times$10$^9$ p/point), i.e. systematic uncertainty, comparing the ML-estimated PE-phantom position with the corresponding known experimental position over the full displacement range for PGI, PET, and the two Compton-imaging channels based on the 511~keV and 718~keV $\gamma$-rays. Section~\ref{sec:reduced-statistics} then examines the robustness of the two best-performing modalities, PGI and PET, when the available event statistics are artificially reduced to $10\%$ and $1\%$ of the nominal dataset, which corresponds to 4$\times$10$^9$p and 4$\times$10$^8$ p. The latter is close to the clinically relevant intensity per spot ($\sim10^{8}$~p) and thus, it will serve to assess the possibility of range-verification in real time as pencil-beam spot intensities range from a few times $10^{7}$ to $10^{9}$ protons depending on the prescribed dose and beam energy~\cite{Ozoemelam:2020b}.


\subsection{Systematic Uncertainty determining the ion-range verification}
\label{sec:full-statistics}

\begin{figure}[htb!]
\centering
\includegraphics[width=0.8\linewidth]{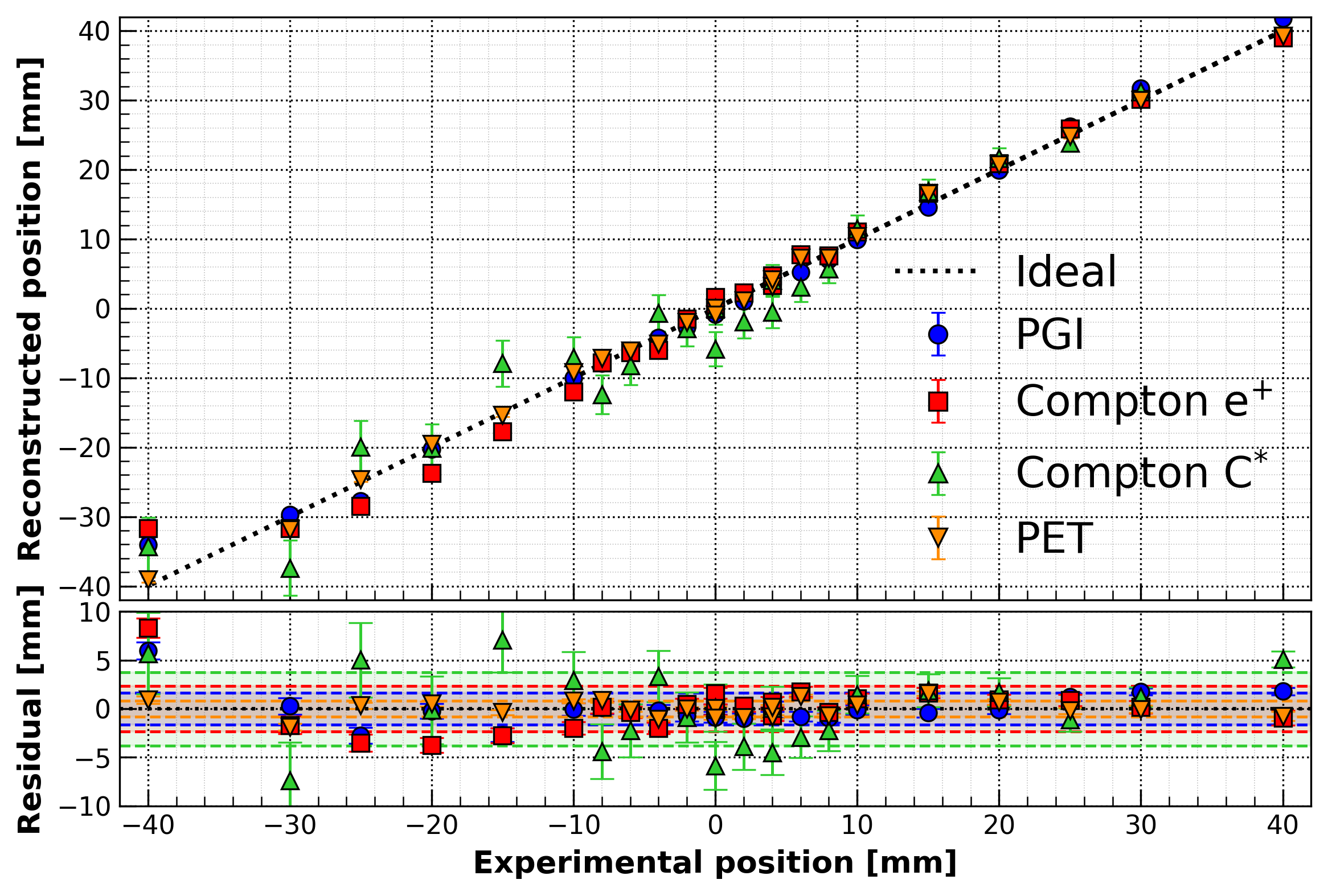}
\caption{Estimated PE-phantom position as a function of the experimentally defined position over the displacement range from $-40$ to $+40$~mm for the four imaging modalities investigated in this work. The upper panel shows the reconstructed positions together with the ideal one-to-one correspondence (black dashed line). The lower panel displays the corresponding residuals, defined as the estimated minus the experimental position. Horizontal dashed lines indicate the mean residual for each modality, while the shaded bands represent the corresponding $\pm1\sigma$ dispersion. Error bars indicate the uncertainties associated with the estimated positions.
}
\label{fig:reconstructed_position_residuals}
\end{figure}

Figure~\ref{fig:reconstructed_position_residuals} summarizes the performance of the four range-verification approaches investigated in this work utilizing 40$\times$10$^9$ p per phantom position. The top panel shows the ML-estimated PE-phantom position as a function of the corresponding experimental position for PGI (blue circles), Compton imaging of e$^{+}$ annihilation photons (red squares), Compton imaging of $^{10}$C$^{*}$ decay (green triangles), and PET (orange inverted triangles). The black dashed line represents the ideal one-to-one correspondence between the estimated and experimental positions. The lower panel shows the corresponding residuals, defined as the estimated minus the experimental position. For each modality, the horizontal shaded band in the bottom panel represents the $\pm1\sigma$ dispersion of the data, which can be regarded as the systematic uncertainty of the methodology for each imaging modality. The black dash-dotted line denotes zero residual. Error bars for each data-point correspond to the uncertainty associated by the error propagation from the linear calibration.

Overall, all four approaches reproduce the expected one-to-one relationship over a broad range of phantom positions, although clear differences in reconstruction accuracy are observed among the different modalities. For PGI, the maximum deviation from the experimental position remains below 4~mm over the interval from $-30$ to $+40$~mm and increases to approximately 6~mm at $-40$~mm. Considering the full displacement range, the residuals yield a root-mean-squared (RMS) deviation of approximately 1.6~mm.

PET provides the best overall performance, exhibiting an approximately linear response over the entire investigated range. The maximum deviation is about 2~mm at an experimental position of $-30$~mm, while the RMS deviation of the residuals is approximately 0.8~mm. Larger discrepancies are obtained for the two Compton-imaging approaches. The reconstruction based on 511~keV annihilation photons yields an RMS deviation of 2.4~mm, with the largest discrepancy reaching approximately 8~mm at $-40$~mm. The reconstruction based on $^{10}$C$^{*}$ decay shows the largest dispersion, with an RMS deviation of 3.8~mm and deviations of several millimeters across the investigated range.

The reduced performance of the two Compton modalities is mainly attributed to their poorer angular resolution and lower counting statistics relative to PGI and, particularly, PET. In particular, the statistics for $^{10}$C$^{*}$ were quite poor. Nevertheless, all four approaches retain sensitivity to changes in the phantom position and follow the expected displacement trend across most of the investigated range.

The RMS deviations, associated with the systematic uncertainty, obtained for the four imaging modalities over the full displacement range are summarized, together with the results of the reduced-statistics study discussed below, in Table~\ref{tab:position_rms}.

It is worth noting that a common systematic deviation is observed for all four modalities at the most negative phantom displacement, suggesting a shared edge effect associated with the reconstruction near the boundary of the field of view rather than a modality-specific bias. In contrast, at $+40$~mm all four approaches retain an accuracy better than 5~mm. This asymmetric behavior arises from the phantom geometry relative to the detector FoV: displacing the phantom towards $-40$~mm progressively moves the relevant emission region out of the FoV, whereas at $+40$~mm it remains more favorably contained within the sensitive region. Thus, entering and exiting the FoV are not geometrically equivalent situations.

\subsection{Sensitivity study for real-time ion-range verification}
\label{sec:reduced-statistics}

\begin{figure}[htbp]
\centering
\includegraphics[width=0.8\linewidth]{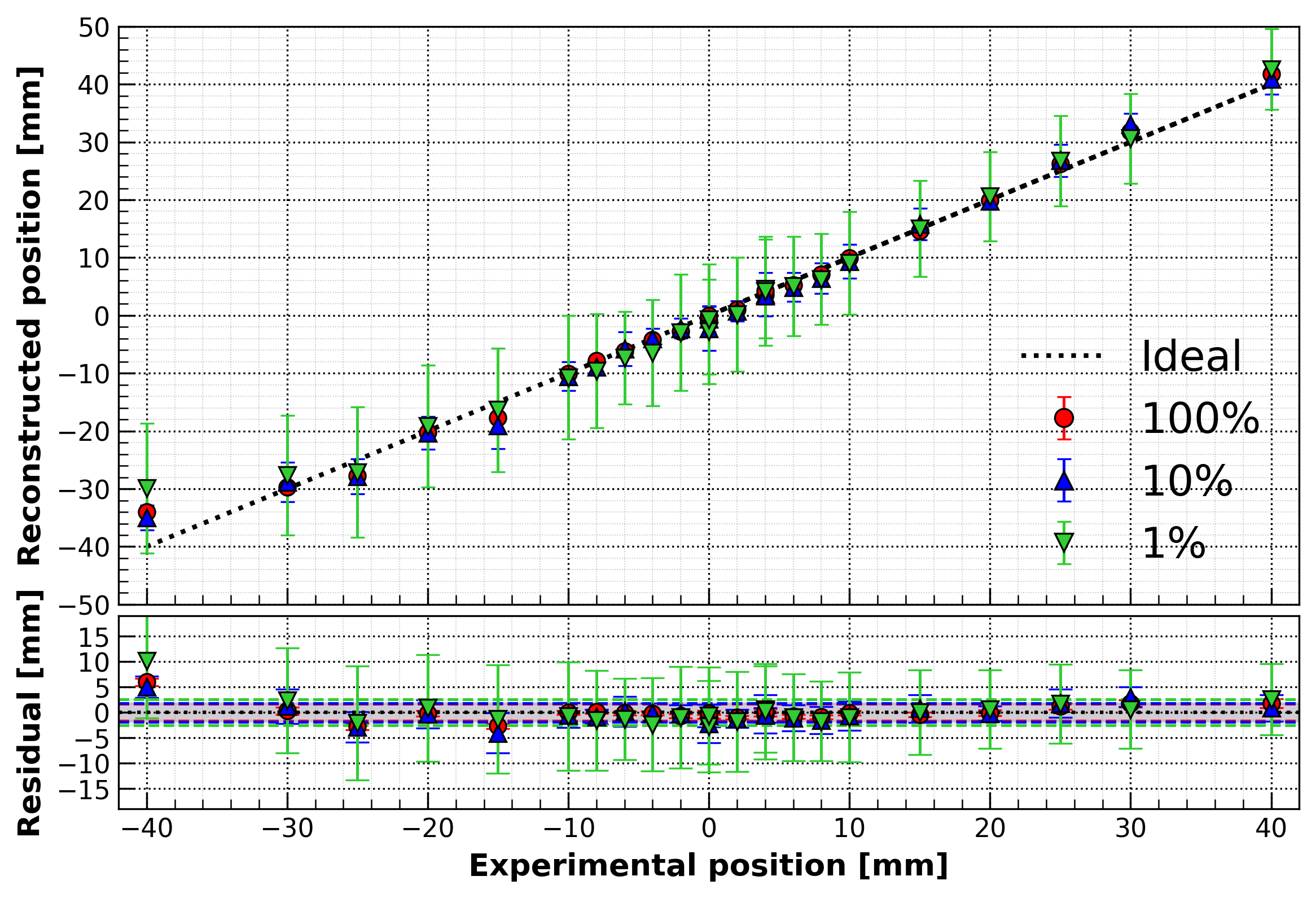}
\caption{Effect of reduced statistics on the PGI position estimation over the phantom displacement range from $-40$ to $+40$~mm. The upper panel shows the estimated phantom position as a function of the experimentally defined position for the nominal dataset and for datasets reduced to $10\%$ and $1\%$ of the nominal statistics. The black dashed line represents the ideal one-to-one correspondence. The lower panel shows the corresponding residuals, defined as the estimated minus the experimental position. Horizontal dashed lines indicate the mean residual for each statistics level, while the shaded bands represent the corresponding $\pm1\sigma$ dispersion. Error bars indicate the uncertainties associated with the individual position estimates.}
\label{fig:compton_sensitivity}
\end{figure}

\begin{figure}[htbp]
\centering
\includegraphics[width=0.8\linewidth]{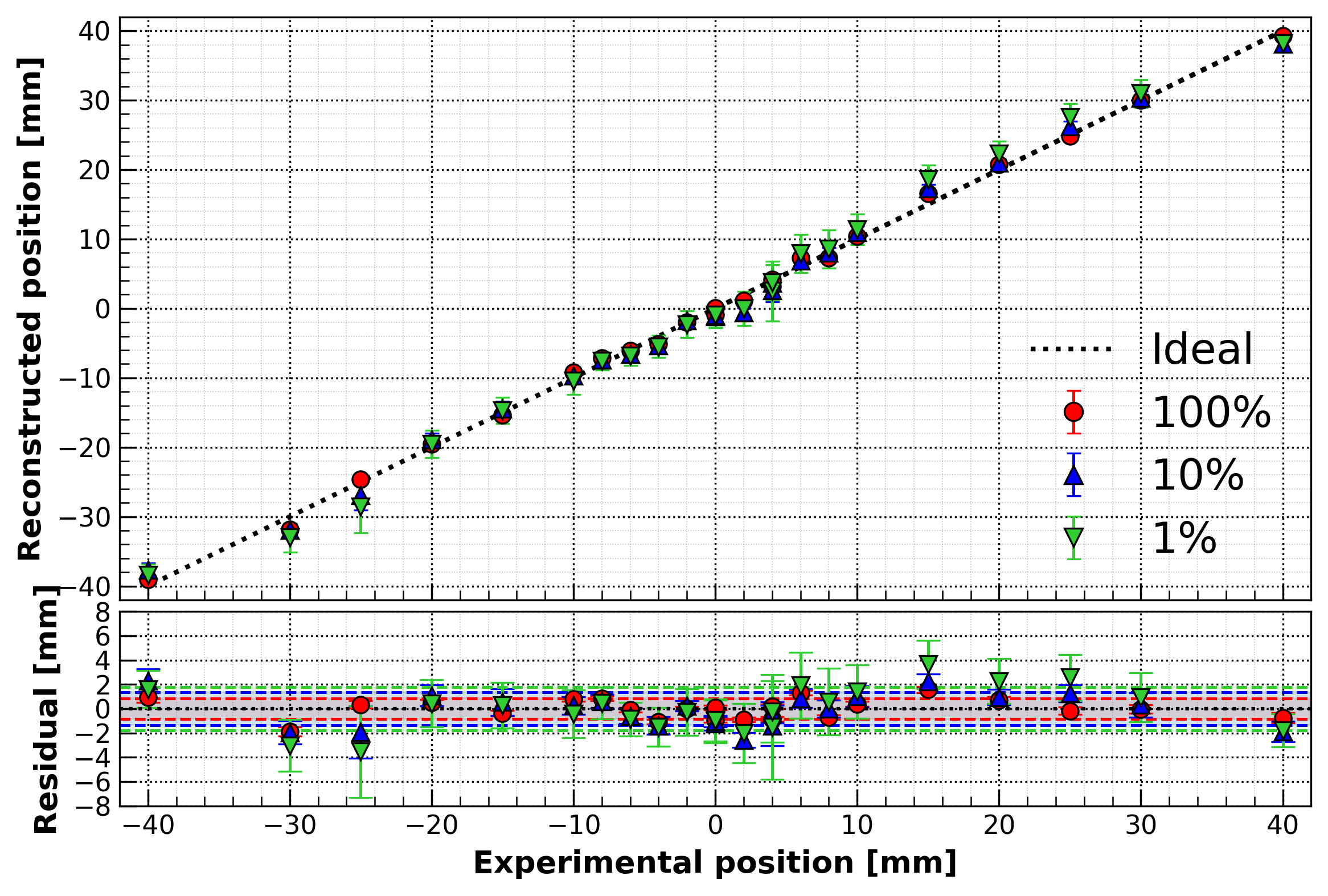}
\caption{Effect of reduced statistics on the PET position estimation over the phantom displacement range from $-40$ to $+40$~mm. The upper panel shows the estimated phantom position as a function of the experimentally defined position for the nominal dataset and for datasets reduced to $10\%$ and $1\%$ of the nominal statistics. The black dashed line represents the ideal one-to-one correspondence. The lower panel shows the corresponding residuals, defined as the estimated minus the experimental position. Horizontal dashed lines indicate the mean residual for each statistics level, while the shaded bands represent the corresponding $\pm1\sigma$ dispersion. Error bars indicate the uncertainties associated with the individual position estimates.}
\label{fig:pet_sensitivity}
\end{figure}

Monitor Units (MU) are used to define the number of protons per spot during treatment. At 100 MeV about 8.1$\times$10$^7$ protons correspond to one MU. A spot of a clinical field can have between 0.025 MU and 10 MU.
Online or real-time ion-range verification requires sufficient detection sensitivity and accuracy at clinically relevant proton-intensities of $\sim$10$^8$ protons per spot, which correspond to $\sim$1.2 MU. Thus, to evaluate the robustness of the ion-range reconstruction under reduced-statistics conditions, the PGI and PET analyses were repeated after artificially reducing the number of available events to $10\%$ (4$\times$10$^9$p/position) and $1\%$  (4$\times$10$^8$p/position) of the nominal dataset statistics. For each statistics level, the full dataset was divided into independent subsets containing the corresponding fraction of events, and the phantom position was estimated separately for each subset. The final position estimate was then obtained as the mean of the individual subset estimates, while their RMS dispersion was taken as the associated statistical uncertainty. The total uncertainty assigned to each position was calculated as the quadratic sum of this statistical contribution and the uncertainty arising from the propagation of the linear-calibration parameters. The resulting position estimates are compared with those obtained using the full dataset in Figs.~\ref{fig:compton_sensitivity} and~\ref{fig:pet_sensitivity} for PGI and PET, respectively.

The mean positions obtained with the $10\%$ and $1\%$ PGI datasets remain compatible with those derived from the full-statistics data for a large fraction of the positions, preserving a good overall linear response across the investigated displacement range, with the exception of the $-40$~mm position as discussed in Section~\ref{sec:full-statistics}. As indicated by the dashed bands in the residual distributions shown in the bottom panel, the systematic uncertainty increases only moderately for PGI, from 1.6~mm for the full-statistics dataset to 1.8~mm and 2.5~mm for the $10\%$ and $1\%$ statistics data sets, respectively. In contrast, the statistical uncertainty shows a much stronger dependence on the number of available events, reaching approximately 1.0~mm for the $10\%$ dataset and no less than 8~mm for the $1\%$ dataset. This substantial increase in statistical uncertainty becomes the dominant limitation at low statistics, severely constraining the sensitivity of the PGI modality for accurate position determination in real time, despite using four Compton modules with large efficiency in rather close geometry. These results indicate that reducing the PGI statistics primarily deteriorates the precision of the position estimate, while introducing only a limited overall systematic bias.


The PET reconstruction exhibits a considerably weaker dependence on the available statistics. The corresponding individual uncertainties are approximately 0.5~mm and 1.5~mm for the $10\%$ and $1\%$ datasets, respectively. RMS deviations are of 1.3~mm, and 1.7~mm, respectively. Although the residual dispersion and the uncertainties of the individual estimates increase as the statistics are reduced, the deterioration is remarkably smaller than for PGI. Even with only $1\%$ of the nominal events, the residuals remain within approximately 2~mm over most of the scanned range, with larger deviations restricted to a limited number of phantom positions.

Table~\ref{tab:position_rms} summarizes the RMS deviations obtained for all four imaging modalities at full statistics, together with those obtained for PGI and PET at $10\%$ and $1\%$ of the nominal statistics, allowing a direct comparison of the accuracy and the statistical robustness of the different approaches. The e$^{+}$  and $^{10}$C$^{*}$ Compton-imaging channels were evaluated only with the full experimental statistics, owing to their already limited event counts at the nominal exposure.

\begin{table}[htb!]
\centering

\begin{tabular}{c c c c c}
\hline
Statistics & PGI & PET & e$^{+}$ & $^{10}$C$^{*}$ \\
(protons) & \textsc{rms} [mm] & \textsc{rms} [mm] &\textsc{rms} [mm] &\textsc{rms} [mm] \\
\hline
4$\times$10$^{10}$ & 1.6 & 0.8 & 2.4 & 3.8 \\
4$\times$10$^{9}$  & 1.8 & 1.3 & --  & --  \\
4$\times$10$^{8}$   & 2.5 & 1.7 & --  & --  \\
\hline
\end{tabular}
\caption{Summary of the RMS deviation, in mm, of the position residuals obtained for each imaging modality over the full displacement range, as a function of the fraction of the nominal event statistics used for the position estimation. The e$^{+}$ and $^{10}$C$^{*}$ Compton-imaging channels were not evaluated at reduced statistics.}
\label{tab:position_rms}
\end{table}

\section{Discussion}
\label{sec:discussion}

The results presented in Section~\ref{sec:full-statistics} show that all four imaging observables investigated in this work exhibit a monotonic and approximately linear response to phantom displacement over most of the scanned range, demonstrating their sensitivity to changes in the proton range. The achievable position accuracy, however, differs substantially among modalities. The RMS deviations summarized in Table~\ref{tab:position_rms} follow the hierarchy PET ($\approx0.8$~mm), PGI ($\approx1.6$~mm), Compton imaging of e$^{+}$ annihilation photons ($\approx2.4$~mm), and Compton imaging of $^{10}$C$^{*}$ decay ($\approx3.8$~mm). These differences arise primarily from the interplay between the available event statistics and the spatial information provided by each detection mechanism. In PET, each coincidence defines a line of response connecting the two detector interactions, thereby providing a strong geometrical constraint on the annihilation position. A Compton event, in contrast, constrains the photon origin to the surface of a cone and therefore carries less spatial information on an event-by-event basis. Consequently, a larger number of Compton events is generally required to achieve a localization precision comparable to PET. The two Compton channels are further limited by their lower event yields. This effect is particularly relevant for the 718~keV channel, for which the comparatively low production yield of $^{10}$C$^{*}$ substantially reduces the available statistics.

The reduced-statistics study presented in Section~\ref{sec:reduced-statistics} further highlights these differences. PET remains remarkably robust as the number of recorded events is reduced, with the RMS deviation increasing only from approximately 0.8~mm for the full dataset to 1.7~mm at $1\%$ of the nominal statistics. PGI shows a stronger dependence on the available statistics. While the dispersion associated with the position reconstruction increases only moderately, from approximately 1.6~mm to 2.5~mm, the statistical uncertainty of the individual position estimates grows substantially, reaching approximately 8~mm at $1\%$ statistics and becoming the dominant contribution under these conditions. This behaviour is consistent with the weaker spatial constraint provided by individual Compton events: as the number of detected events decreases, the reconstructed emission profile becomes progressively less well constrained, and substantially more statistics are required for the Compton-cone ensemble to converge towards a stable spatial distribution. These experimental observations are consistent with the design considerations discussed in Ref.~\cite{Krimmer:18}, which emphasize the importance of maximizing the detection efficiency and angular coverage of Compton-camera systems for real-time ion-range verification. In this respect, the multi-module i-TED configuration employed here represents an important step towards obtaining the statistics required for clinically relevant PGI measurements.

A useful comparison can be made with dedicated studies of $^{12}$N-based quasi-prompt PET range verification in proton therapy. For pencil-beam spots containing approximately $10^{8}$ protons, range precisions of 2.5--2.6~mm have been reported, with sub-2~mm precision projected within the first 50~ms of irradiation for optimized detector configurations~\cite{Ozoemelam:2020b}. At a comparable statistics level, the present approach yields RMS deviations of approximately 2.5~mm for PGI and 1.7~mm for PET. Although the reconstruction methodologies and experimental conditions are not identical, these values are comparable to clinically oriented benchmarks and support the feasibility of the hybrid PGI--PET concept for range verification at realistic single-spot intensities.

The different temporal characteristics of PGI and PET are equally important when considering their practical use. PET provides the highest position-estimation accuracy among the observables investigated here, but its signal is intrinsically delayed because it relies on the accumulation of $\beta^{+}$ decays during the beam-off periods of the PBS delivery sequence. Over longer time scales, the reconstructed activity distribution may also be affected by biological washout in a clinical environment. PGI, by contrast, provides a prompt signature correlated with the proton interaction region during beam delivery. Despite its somewhat lower accuracy and stronger statistical dependence, this makes PGI particularly attractive for rapid range monitoring and for identifying large deviations as early as possible during irradiation. The two modalities should therefore be regarded as complementary rather than competing: PGI provides prompt information, whereas PET offers a more precise but delayed estimate that can subsequently refine or confirm the PGI-based result once sufficient step period statistics have accumulated. The present measurements are consistent with, and provide quantitative support for, this complementary picture under beam-delivery conditions representative of clinical proton therapy.

The common systematic deviation observed for all four modalities at the most negative phantom displacement also illustrates an important geometrical limitation of the present setup. In particular, the PET field of view is constrained by the dimensions of the absorber planes to approximately $120\times120$~mm$^2$. As the relevant emission region approaches and progressively leaves this sensitive volume, edge effects become increasingly important and lead to biased position estimates. The asymmetry observed between positive and negative phantom displacements reflects the fact that entering and leaving the effective FoV are not geometrically equivalent for the present detector configuration. Extending the usable FoV could be achieved by increasing the detector-plane dimensions, modifying the module separation, increasing the number of detector modules, or adopting a more complete ring-like geometry. Such modifications, however, must be balanced against the requirements of PGI, for which the compact, co-planar, cross-shaped arrangement provides high solid-angle coverage and enhanced detection efficiency compared with the twofold face-to-face geometry employed in the previous HIT campaign~\cite{Balibrea:2025a}. Determining the optimal compromise between PET FoV and PGI efficiency therefore represents an important aspect of future detector optimization.

Several limitations of the present study should also be considered. Residual discrepancies between the experimental and MC-simulated one-dimensional profiles, shown in Fig.~\ref{fig:FitDistributions}, are most evident for the step observables and are attributed primarily to the simplified description of the relative production yields of the different $\beta^{+}$-emitting isotopes in the simulations. A more realistic modeling, or experimental tuning, of the isotope-production yields could improve the agreement between simulated and measured PET profiles and consequently reduce the systematic corrections required by the ML-based position-estimation procedure. The PGI distributions, which are not affected by this particular source of uncertainty, showed a closer agreement with the corresponding simulations.

Detector dead-time and pile-up effects were also not included in the MC datasets used for training. These effects are expected to be particularly relevant for the high instantaneous counting rates encountered during PGI acquisition and will require dedicated treatment in future studies. Furthermore, the present campaign was restricted to a proton energy of 100~MeV and to a single phantom material. Previous measurements at HIT showed a significant dependence of both PGI and PET observables on beam energy and ion species~\cite{Balibrea:2025a}; consequently, the present results should not be extrapolated directly to other clinically relevant beam energies, ion species, or patient-like geometries without further experimental validation. Finally, the ML analysis considered each imaging modality independently. A natural extension of the present methodology would be to combine the information from several, or all, reconstructed profiles within a single multimodal regression model. Such an approach could exploit the complementary spatial and temporal information provided by PGI and PET and may further improve both the accuracy and robustness of the range-verification procedure. Beyond the broadband PGI signal exploited in the present work, the discrete characteristic $\gamma$-ray lines introduced in Section~\ref{sec:introduction} (\textit{e.g.} the 4.44~MeV and 6.13~MeV transitions) could, in principle, be isolated via spectral deconvolution~\cite{Jeyasugiththan:21,HOSOBUCHI:23} to provide an additional, composition-sensitive range-verification observable, a direction we intend to explore in future work.

A related simplification concerns the dimensionality of the position-estimation problem addressed here. Although the PET and Compton-PGI activity distributions were reconstructed in three dimensions (Figs.~\ref{fig:compton3d_distribution} and~\ref{fig:pet3d_distribution}), the ML models were trained and applied exclusively to their one-dimensional projections along the beam axis, discarding the transverse spatial information contained in the full reconstruction. While this choice was adequate for the essentially one-dimensional displacement scans performed in the present campaign, an actual clinical range-verification scenario involves genuinely three-dimensional anatomical and dosimetric uncertainties. Exploiting the full three-dimensional reconstructed distributions, rather than their axial projections, within the ML-based position-estimation framework therefore represents an important extension of the present methodology.


\section{Conclusions and future work}
\label{sec:conclusions}

We performed a comprehensive experimental investigation of a hybrid PGI--PET range-verification system using four i-TED Compton-camera modules arranged in a cross-shaped configuration, under beam-delivery conditions representative of clinical pencil-beam scanning treatments at the West German Proton Therapy Centre. Exploiting the natural alternation between shoot and step periods of the irradiation sequence, the position of polyethylene phantoms irradiated with 100~MeV protons was estimated from one-dimensional PGI, PET, and Compton profiles using a supervised machine-learning model trained exclusively on MC simulations and applied directly to the experimental data. All four modalities showed a monotonic, approximately linear response to phantom displacement, with PET achieving the best accuracy (RMS~$\approx0.8$~mm), followed by PGI ($\approx1.6$~mm) and the e$^{+}$ and $^{10}$C$^{*}$ Compton channels ($\approx2.4$ and $3.8$~mm, respectively). PET was also the most statistically robust modality, retaining an RMS deviation of approximately 1.7~mm at only $4\times10^{8}$ protons, whereas PGI became increasingly limited by statistical uncertainty at reduced event counts.

These results are consistent with the expected complementarity of the hybrid concept: PGI offers a prompt signature well suited to rapid range monitoring, while PET provides a more accurate, statistically robust estimate once sufficient step-period statistics have accumulated. The systematic deviations observed near the edge of the field of view underline the need for further detector-geometry optimization, particularly for PET. The present one-dimensional ML analysis, applied only to axial projections of the fully three-dimensional PET and Compton-PGI reconstructions (Figs.~\ref{fig:compton3d_distribution} and~\ref{fig:pet3d_distribution}), will be extended to the full 3D volumes in future work, together with improved MC modelling of $\beta^{+}$-emitter yields and dead-time/pile-up effects, measurements at additional energies, ion species, and phantom geometries, and a multimodal ML estimator combining all four observables.

A particularly promising direction for this future work is the extension of the present methodology to IBA Proteus ONE facilities, which employ a superconducting synchrocyclotron delivering an intrinsically pulsed beam, with spot durations of approximately 7--10\,$\mu$s and inter-pulse gaps of the order of 1\,ms~\cite{Pidikiti18}. Unlike the isochronous cyclotron used at WPE, where the shoot/step structure had to be imposed for the purposes of the present study, this time structure is naturally available in Proteus ONE systems, closely resembling, on a much shorter time scale, the shoot and step periods exploited here. Implementing the present hybrid PGI--PET approach on such a synchrocyclotron would therefore allow the methodology to be validated under a genuinely native clinical beam-delivery scheme, without requiring any imposed modification of the accelerator time structure, and represents a natural next step towards its clinical translation.

Altogether, this campaign extends previous proof-of-concept and preclinical studies to clinically representative beam-delivery conditions and supports the continued development of hybrid PGI--PET imaging towards reliable, millimetre-scale proton-range verification, as a step towards its eventual \textit{in vivo} clinical application.

\ack{We thank S\'everine Rossomme and Julien Smeets (IBA) for insightful discussions and constructive feedback throughout this work. The corresponding author acknowledges the use of the generative AI tool Claude (Sonnet 5, Anthropic) and Nature research assistant to assist in revising the wording of selected passages of the manuscript, including clarifying technical terminology, moderating overstated claims, and improving sentence flow. The AI tool was not used for data analysis, generation of scientific content or results, or interpretation of the findings; all suggested edits were reviewed, verified, and approved by the authors, who take full responsibility for the content of the manuscript.
}

\funding{This work derives from fundamental research carried out with funding from the European Research Council (ERC) under the European Union's Horizon 2020 research and innovation programme (ERC Consolidator Grant project HYMNS, with grant agreement No.~681740) and part of this work has been funded under ERC Proof-of-Concept Grant GNVISION (Grant Agreement 101113330). The authors acknowledge support from the Spanish Ministerio de Ciencia e Innovación under grants PDC2021-121536-C21, PID2022-138297NB-C21. We acknowledge support from the Severo Ochoa Grant CEX2023-001292-S funded by MCIU/AEI.}

\roles{
JB-C: conceptualization, methodology, software, formal analysis, investigation, data curation, visualization, writing -- original draft.
VB-S: investigation, software, data curation.
JL-M: methodology, software, investigation, writing -- review \& editing.
PT-S: investigation, data curation.
IL: hardware, software, visualization.
MP: software, formal analysis.
BB: software, formal analysis.
JW: resources, investigation, supervision.
CB: resources, supervision, investigation.
AT-S: investigation, methodology.
CD-P: conceptualization, methodology, supervision, project administration, funding acquisition, writing -- review \& editing.}

\data{Data will be available on reasonable request.}


 \bibliographystyle{unsrt}
 \bibliography{sample}

\end{document}